\documentclass[useAMS, onecolumn]{mn2e}
\usepackage{times}
\usepackage{rotating}
\long\def\symbolfootnote[#1]#2{\begingroup%
\def\thefootnote{\fnsymbol{footnote}}\footnote[#1]{#2}\endgroup}
\newcommand{\gae}{\lower 2pt \hbox{$\, \buildrel {\scriptstyle >}\over {\scriptstyle
\sim}\,$}}
\newcommand{\lae}{\lower 2pt \hbox{$\, \buildrel {\scriptstyle <}\over {\scriptstyle
\sim}\,$}}

\begin{document}
 
\title[Inverse Compton in Klein-Nishina regime in GRBs]{Inverse Compton cooling 
in Klein-Nishina regime and GRB prompt spectrum}

\author[Barniol Duran, Bo\v{s}njak \& Kumar]{R. Barniol Duran$^1$\thanks
{E-mail: rbarniol@phys.huji.ac.il, zeljka.bosnjak@cea.fr, pk@astro.as.utexas.edu},
\v{Z}. Bo\v{s}njak$^2$\footnotemark[1] and P. Kumar$^3$\footnotemark[1] \\
$^{1}$Racah Institute of Physics, Edmund J. Safra Campus, Hebrew University of Jerusalem, Jerusalem 91904, Israel \\
$^{2}$AIM (UMR 7158 CEA/DSM-CNRS-Universit\' e Paris Diderot) Irfu/Service d'Astrophysique,  Saclay, 91191 Gif-sur-Yvette Cedex, France\\
$^{3}$Department of Astronomy, University of Texas at Austin, Austin, TX 78712, USA} 

\date{Accepted; Received; in original form 2011 December 22}

\pubyear{2012}

\maketitle

\begin{abstract}
Synchrotron radiation mechanism, when electrons are accelerated
in a relativistic shock, is known to have serious problems 
to explain the observed gamma-ray spectrum below the peak 
for most Gamma-Ray Bursts (GRBs); the synchrotron spectrum 
below the peak is much softer than observed spectra. Recently, the
possibility that electrons responsible for the radiation cool 
via Inverse Compton, but in the Klein-Nishina  regime, has been 
proposed as a solution to this problem. We provide an analytical 
study of this effect and show that it leads to a hardening of
the low energy spectrum but not by enough to make it consistent
with the observed spectra for most GRBs (this is assuming that
electrons are injected continuously over a time scale comparable 
to the dynamical time scale, as is expected for internal shocks
of GRBs). In particular, we find that it is not possible to 
obtain a spectrum with $\alpha>-0.1$ ($f_{\nu} \propto 
\nu^{\alpha}$) whereas the typical observed value is $\alpha\sim0$.
 Moreover, extreme values for a number of parameters are 
required in order that $\alpha\sim -0.1$: the energy fraction 
in magnetic field needs to be less than about $10^{-4}$, the 
thermal Lorentz factor of electrons should be larger than 10$^6$, 
and the radius where gamma-rays are produced should be 
not too far away from the deceleration radius.
These difficulties suggest that the synchrotron radiation mechanism 
in internal shocks does not provide a self-consistent solution when 
$\alpha \gae -0.2$.

\end{abstract}


\section{Introduction}

The dissipation mechanism responsible for the prompt emission of 
Gamma-Ray Bursts (GRBs) remains unknown.  There have been various 
ideas put forth to explain it (for a review, see Piran 1999, 2004; M\'esz\'aros
2006; Gehrels et al. 2009). One of the main problems is to explain the fact 
that the majority of GRBs exhibit a spectrum $f_{\nu} \propto \nu^{\alpha}$, 
with $\alpha \sim 0$ below the peak of the spectrum (Preece et al. 2000), 
whereas the simplest version of the synchrotron model 
(in the so-called ``fast cooling regime'') predicts $\alpha=-1/2$ 
(see, e.g., Ghisellini et al. 2000).  Recently, a modified version of the 
synchrotron model, in which electrons that radiate below the peak of the 
spectrum cool via Inverse Compton (IC) in the Klein-Nishina (KN) regime 
(Derishev et al. 2003) has gained popularity (Bo\v{s}njak et al. 2009, 
Nakar et al. 2009, Wang et al. 2009, Fan 2010, Daigne et al. 2011).  The 
main idea is very simple: Electrons cooling via synchrotron mechanism (or 
IC in the Thomson regime) exhibit an energy loss rate $\propto
\gamma_e^{\delta}$, where $\gamma_e$ is the electron Lorentz Factor (LF) and 
$\delta=2$.  The observed synchrotron spectrum is then 
$f_{\nu} \propto \nu^{-(\delta-1)/2} = \nu^{-1/2}$.  However, when the cooling 
of electrons is dominated by the IC in the KN regime, where the
photon-electron interaction cross section scales as $\sim \gamma_e^{-1}$, 
then $\delta \approx 1$ and $f_{\nu} \sim \nu^{0}$. In this paper, 
we investigate this scenario in detail and explore its consequences. 

We analytically study the IC cooling in the KN regime assuming 
that electrons are injected continuously over a time scale comparable 
to the dynamical time scale, as is expected for the internal shock model 
of GRBs (Piran, Shemi \& Narayan 1993; Katz 1994; Rees \& M\'esz\'aros 1994).  
Recently, Daigne et al. (2011) have provided a detailed numerical calculation 
of the same context; however, they have assumed that electrons are injected 
instantaneously in the internal shock.  Therefore, our work and the work of 
Daigne et al. (2011) are complementary. Essentially, Daigne et al. (2011)
deals with the case when electrons are no longer being injected and they 
simply cool, which happens when the internal shock has already passed through 
the shell. They consider the superposition of emission of many shells, for
which the shock has crossed all of them, and the shells simply adiabatically 
expand and cool.  We, however, consider the shock as it traverses the shell, 
accelerates electrons, and these radiate.

The scenario presented in this paper has been considered before 
(Nakar et al. 2009; Fan 2010) and our results are consistent.  However, in 
contrast with these works and with the work of Daigne et al. (2011), the work 
presented here is applied to the prompt phase data of particular GRBs with 
$\alpha \approx 0$: we analyze the data of GRB 080916C, a burst detected by
the {\it Fermi} Satellite, in the context of the scenario described above, and 
provide constraints on this scenario based on available $>$100 MeV data.

Recent developments on prompt GRB theory have cast doubt on the internal 
shock model (see, e.g., Kumar \& Narayan 2009, Zou et al. 2009).  New
alternative models have been proposed to solve the low-energy spectral index
problem described above and other prompt theory issues (see, e.g.,
M\'esz\'aros \& Rees 2000, Drenkhahn \& Spruit 2002; Lyutikov \& Blandford
2003; Giannios 2008; Narayan \& Kumar 2009; Kumar \& Narayan 2009; Lazar et al. 2009; 
Beloborodov 2010; Vurm et al. 2011; M\'esz\'aros \& Rees 2011; Ioka 2010; Ioka
et al. 2011; Zhang \& Yan 2011, Bo\v{s}njak \& Kumar 2012; Pe'er et al. 2012).  
It is, however, still relevant to critically test the internal shock model, in
the particular case where electrons cool via IC in the KN regime, to  assess
its feasibility.

It is important to mention that there exists a fraction of GRBs with 
$0 < \alpha < 1/3$ (only about 25 per cent of GRBs have more than 50 per cent 
of their spectra with $\alpha$ in this range; see Kaneko et al. 2006), that
is, with spectra consistent with synchrotron radiation mechanism; however the 
scenario presented here is unable to explain them.  In this work we focus on
the {\it majority} of GRBs, which have $\alpha \approx 0$; in particular, we
study the case of GRB 080916C, which shows $\alpha = -0.02 \pm 0.02$ for most 
of its duration (Abdo et al. 2009).

We set up our model and present the relevant time scales in Section 2.  
In Sections 3 and 4, we calculate the effect of IC cooling in the KN regime on
the electron energy distribution and on the observed spectral slope,
respectively.  In Section 5 we derive the relevant physical parameters (radius
of emission and total luminosity).  In Section 6 we apply our results to
GRB080916C, and to an average long-duration GRB.  In Sections 7 and 8, we
present a Discussion and our Conclusions.

\section{Electron cooling}

Let us consider a GRB jet that has bulk LF $\Gamma$ and bolometric 
$\gamma$-ray luminosity (isotropic equivalent) $L$. The peak of the GRB spectrum
($\nu f_{\nu}$) in observer frame is $\nu_p$. We define

\begin{equation}
\epsilon_p \equiv \frac{h \nu_p (1+z)}{m_e c^2},
\end{equation}
where $z$ is the redshift and $h$, $m_e$ and $c$ are Planck's constant, 
the electron mass and the speed of light, respectively.

The case we are trying to explain using the idea of electron cooling via
IC cooling in the KN regime is when the observed spectrum below the peak 
is $f_{\nu} \propto \nu^{\alpha}$, with $\alpha \sim 0$; $\alpha$ is
known as the low energy spectral index. We take $\alpha$ to extend from 
at least 10 keV to $\nu_p$; 10 keV is roughly the lower energy limit of 
the GBM detector on board the {\it Fermi} satellite.

Let us take the thermal LF of electrons that produce 10 keV photons (via synchrotron process) 
to be $\gamma_4$, and the LF of electrons producing photons of frequency $\nu_p$ to be $\gamma_i$.
We start with the assumption that electrons with LF $\gamma_4<\gamma_e<\gamma_i$ cool primarily via 
IC in the KN regime.  In order to satisfy this assumption, 
$\gamma_4$ should be such that 

\begin{equation} \label{KN_condition}
\frac{\gamma_4(1+z)h \nu_p}{\Gamma}>m_ec^2,
\end{equation} 
which leads to the condition that $\gamma_4>\epsilon_p^{-1}\Gamma$. We define 
a variable $\eta_4$ ($\eta_4 \geq 1$) which tells us how deep electrons
of $\gamma_4$ are in the KN regime.  With this, $\gamma_4$ is

\begin{equation} \label{gamma_4}
\gamma_4=\eta_4 \epsilon_p^{-1}\Gamma.
\end{equation}

We can find the magnetic field strength in the jet comoving frame, $B$, so that electrons with LF 
$\gamma_4$ have synchrotron radiation at 10 keV in the observer frame.  The 
observed synchrotron frequency of electrons of $\gamma_4$ is

\begin{equation}
\nu=\frac{e B\gamma_4^2\Gamma}{2\pi m_ec(1+z)}=(1.15\times10^{-8}\rm eV)\frac{B\gamma_4^2\Gamma}{(1+z)},
\end{equation}
where $e$ is the electron charge. For $\nu=10$ keV, and using (\ref{gamma_4}),  
we find the magnetic field

\begin{equation} \label{B_field}
B=(8.7\times10^2 \rm G)(1+z)\epsilon_p^2\eta_4^{-2}\Gamma_3^{-3},
\end{equation}
where here and throughout the paper we use the usual notation 
$Q_n = Q/10^n$, with the exception of $\gamma_e$, $\eta$
and the Compton-$Y$ parameter, $Y$; in these cases the subscript 
indicates the log$_{10}$ of the observed synchrotron frequency in eV
 we are referring to.

The cooling time due to synchrotron radiation for an electron of 
LF $\gamma_e$, in the jet comoving frame, 
is given by 

\begin{equation} \label{t_syn}
t'_{syn}=\frac{6\pi m_e c}{\sigma_T B^2 \gamma_e}=(7.7\times10^8 \rm s)B^{-2}\gamma_e^{-1},
\end{equation}
where $\sigma_T$ is the Thomson cross section.  For $\gamma_e=\gamma_4$  
we find, using (\ref{gamma_4}) and (\ref{B_field}), that

\begin{equation} \label{t_syn_gamma_4}
t'_{syn}=(1.1s)\frac{\eta_4^3\Gamma_3^5}{(1+z)^2\epsilon_p^3}.
\end{equation}

We now calculate the electron cooling time due to IC 
scattering of $\gamma$-ray photons.  The cross section for scattering photons
of frequency $\nu_p$ by electrons of $\gamma_4$ 
is smaller than the Thomson cross section by a factor of $\approx \eta_4$. Thus,

\begin{equation} \label{IC_KN}
t'_{IC} \approx \frac{4\pi R^2\Gamma^2 m_e c^2}{\gamma_4 (\sigma_T L/\eta_4)}=(154 \rm s)\frac{R_{15}^2\Gamma^2_3\eta_4}{\gamma_4 L_{53}},
\end{equation}
or by substituting (\ref{gamma_4}), we find

\begin{equation}
t'_{IC} \approx (0.15 \rm s) R_{15}^2 \Gamma_3 \epsilon_p L_{53}^{-1}.
\end{equation}
Note that the IC
cooling time is essentially independent of $\gamma_e$ in the KN regime.  The reason is 
that the energy of electrons is $m_e c^2 \gamma_e$ and 
the IC power in the KN regime is approximately 
$\propto \gamma_e$, therefore, the time scale is almost 
independent of $\gamma_e$.  We will compare these cooling time scales 
with the dynamical time in the jet comoving frame

\begin{equation} \label{t_dyn}
t'_{dyn}=\frac{R}{c \Gamma}=(33 \rm s)R_{15}\Gamma_3^{-1}.
\end{equation}

\section{Effect of IC cooling in KN regime on electron distribution}

The electron energy distribution, $n_e$, in steady state, for electrons 
of LF $\gamma_e$, is determined from the continuity equation 

\begin{equation} \label{eq:continuity}
\frac{\partial}{\partial \gamma_e} 
\left[ \dot{\gamma_e} n(\gamma_e) \right] = S(\gamma_e) \propto \left\{ \begin{array}{ll}
\left(\frac{\gamma_e}{\gamma_i}  \right)^{-p}  & \textrm{$\gamma_e \ge \gamma_i$} \\
0 & \textrm{$\gamma_e < \gamma_i$}, \\
\end{array} \right.,
\end{equation}
and the cooling of electrons is determined by

\begin{equation} \label{electron_cooling}
- \dot{\gamma}_e =  \frac{\sigma_T B^2 \gamma_e^2}{6 \pi m_e c} + \frac{\sigma_{KN} L \gamma_e^2}{4 \pi R^2 \Gamma^2 m_e c^2},
\end{equation}
where $\sigma_{KN}$ is the  KN cross section which we write as
$\sigma_{KN}=\sigma_T f(\eta)$, and

\begin{equation} \label{sigma_KN}
f(\eta) = \frac{3}{4}\left[\frac{1+\eta}{\eta^3}\left(\frac{2\eta(1+\eta)}{1+2\eta} 
    - \ln(1+2\eta)\right)+\frac{\ln(1+2\eta)}{2\eta}-\frac{1+3\eta}{(1+2\eta)^2} \right].
\end{equation}
We also define $\eta$ as $\eta=\epsilon_p \gamma_e/\Gamma$ 
analogous to (\ref{gamma_4}), and it indicates how deep  
electrons of $\gamma_e$ are in the KN regime, defined in (\ref{KN_condition}).
In this section we keep the dependence of $\eta$ on $\gamma_e$. 
As a reminder, $\eta$ was defined for a specific
$\gamma_e=\gamma_4$ in the previous section; we will return to 
that same definition later on. With the use of (\ref{sigma_KN}), 
equation (\ref{electron_cooling}) can be rewritten as 

\begin{equation}
- \dot{\gamma}_e =  \frac{\gamma_e^2}{T'_{syn}} + \frac{\gamma_e^2}{T'_{IC}}f(\eta),
\end{equation}
where $T'_{syn}$ and $T'_{IC}$ are defined as 

\begin{equation}
T'_{syn} \equiv \frac{6 \pi m_e c}{\sigma_T B^2},
\end{equation}
and

\begin{equation}
T'_{IC} \equiv \frac{4 \pi R^2 \Gamma^2 m_e c^2}{\sigma_T L}.
\end{equation}
Note that $t'_{syn}=T'_{syn}/\gamma_e$ is the synchrotron 
cooling time, defined in (\ref{t_syn}), and 
$t'_{IC,KN}=T'_{IC}\sigma_T/(\sigma_{KN}\gamma_e)=T'_{IC}/(f(\eta) \gamma_e)$ is 
the IC cooling time in the KN regime (see eq. (\ref{IC_KN}) and 
note that $f(\eta)\sim \eta^{-1}$ for $\eta \gg 1$).  

We can define  $Y_{KN} \equiv t'_{syn}/t'_{IC,KN} =
T'_{syn}/(T'_{IC}/f(\eta))$, 
and identify it as the Compton-$Y$ parameter in the KN
regime for electrons with LF $\gamma_e$, which are deep in the KN regime as 
characterized by their parameter $\eta$. The electron cooling rate is 
now 

\begin{equation} \label{electron_cooling2}
- \dot{\gamma}_e =  \frac{\gamma_e^2}{T'_{syn}}\left(1 + Y_{KN}\right).
\end{equation}

Since we are interested in the flux below the peak, we consider electrons 
with LF $\gamma_e < \gamma_i$.  Moreover, we need to consider the case 
$\gamma_c < \gamma_e < \gamma_i$, where $\gamma_c$ is the cooling LF, 
which is the LF of electrons that cool on a dynamical time, $t_{dyn}$.  The 
case $\gamma_e < \gamma_i < \gamma_c$ gives $f_{\nu} \propto \nu^{1/3}$ 
below the peak, which is impossible to obtain for electrons accelerated in
shocks for GRBs (Ghisellini et al. 2000; Kumar \& McMahon 2008).

The solution of the continuity equation (\ref{eq:continuity}) for the the
electron distribution for $\gamma_e<\gamma_i$ is $n_e\propto{\dot{\gamma_e}}^{-1}$ 
and, using (\ref{electron_cooling2}), the corresponding power-law index of the
distribution is


\begin{equation} \label{eq2:p_1}
p_1 \equiv \left|\frac{d \ln n_e}{d \ln \gamma_e}\right| = 2 + \frac{Y_{KN}\frac{d f(\eta)}{d \eta} \eta  }{f(\eta)(1 + Y_{KN})}.
\end{equation}
As mentioned before, $\eta$ depends on $\gamma_e$, and using 
equation (\ref{sigma_KN}) we find

\begin{equation} \label{df}
\frac{df(\eta)}{d\eta} = \frac{3}{4} \left[\frac{2\eta(2\eta^4-39\eta^3-63\eta^2-34\eta-6)-(2\eta+1)^3(\eta^2-4\eta-6)\ln(2\eta+1)}{2\eta^4(2\eta+1)^3} \right].
\end{equation}

For $\eta \ge 1$, which we consider here, eq. (\ref{df}) always yields a
negative value, and since $Y_{KN} \ge 0$, then $1 \le p_1 \le 2$. 
Since we are interested in $f_{\nu} \propto \nu^0$, we will focus 
on finding the corresponding $\eta$ and $Y_{KN}$ that give 
$p_1\rightarrow 1$, so that $\alpha \rightarrow 0$.
Let us first calculate $\alpha$ for a given $p_1$.

\section{Spectral slope when electron distribution is close to $\gamma_e^{-1}$}

Let us consider the electron energy distribution to be 

\begin{equation} \label{eq:ne2}
n_e \propto  \left\{ \begin{array}{ll}
\left(\frac{\gamma_e}{\gamma_i}\right)^{-p_1} & \textrm{$\gamma_e < \gamma_i$} \\
\left(\frac{\gamma_e}{\gamma_i}\right)^{-p_2} & \textrm{$\gamma_e > \gamma_i$}. \\
\end{array} \right.
\end{equation}
We will take $p_1$ to be very close to 1 and $p_2 \approx 2$. The exact value 
of $p_1$ can be found with (\ref{eq2:p_1}).

The synchrotron flux is given by 

\begin{equation} \label{eq:syn}
f_{\nu_o} = A \int_{\gamma_{\nu_o}}^{\infty} \! d\gamma_e  n_e \left[ \frac{\nu_o}{\nu(\gamma_e)}\right]^{1/3}, 
\end{equation}
where $A$ is a constant proportional to $B$, and $\gamma_{\nu_o}$ is the LF of 
electrons radiating at synchrotron frequency $\nu_o$, which is

\begin{equation}
\gamma_{\nu_o} = \left(\frac{2 \pi m_e c \nu_o}{e B}\right)^{1/2}.
\end{equation}
The frequency $\nu(\gamma_e)$ is the synchrotron frequency of electrons with 
LF $\gamma_e$

\begin{equation}
\nu(\gamma_e)=\frac{e B \gamma_e^2}{2 \pi m_e c}.
\end{equation}
Substituting the last two expressions into (\ref{eq:syn}), we can 
integrate this expression using (\ref{eq:ne2}).  By absorbing 
constants in a new variable $A'$, we find

\begin{equation} 
f_{\nu_o} = A' \nu_o^{1/3} \frac{\gamma_i^{1/3}}{p_1 - \frac{1}{3}} \left[\left(\frac{\gamma_{\nu_o}}{\gamma_i}\right)^{-p_1+1/3} - 
\frac{p_2 - p_1}{p_2 - \frac{1}{3}} \right]. 
\end{equation}
With this, we can find the spectral index for synchrotron radiation to be

\begin{equation} \label{eq:alpha}
 \alpha \equiv \frac{d \ln f_{\nu_o}}{d \ln \nu_o} = \frac{1}{3} - \frac{\frac{1}{2}\left(p_1 - \frac{1}{3} \right)}{1 - \frac{p_2 - p_1}{p_2 - 
\frac{1}{3}} \left(\frac{\gamma_{\nu_o}}{\gamma_i}\right)^{p_1-1/3}  } . 
\end{equation}
Since we are interested in the low energy spectral index, $\alpha$, at $\nu_o=10$ keV, 
$\frac{\gamma_{\nu_o}}{\gamma_i}=\left(\frac{10 \rm keV}{\nu_p}\right)^{1/2}$.
For example, for $\nu_p=1$ MeV, then
$\frac{\gamma_{\nu_o}}{\gamma_i}=0.1$, and for $p_1 = (1.02,1.1,1.15,1.3)$ and
$p_2 =2$ we find $\alpha = (-0.057,-0.089,-0.109,-0.173)$.
Note that from (\ref{eq:alpha}) $\alpha<0$ for $p_1\ge1$.

Since we want the spectrum at 10 keV to have $\alpha \approx 0$, we will choose 
the LF of electrons to be $\gamma_e = \gamma_4$.  We will determine how deep 
these electrons have to be in the KN regime ($\eta_4$) and their Compton-$Y$ 
parameter, $Y_{KN,4}$, so that we can obtain $p_1 \rightarrow 1$.  We describe 
the calculation in the next section.    

\section{Physical parameters consistent with $\alpha \rightarrow 0$}

\begin{figure*}
\begin{center}
\includegraphics[width=12cm, angle = 0, clip=true, viewport=.0in .0in 8in 4.5in]{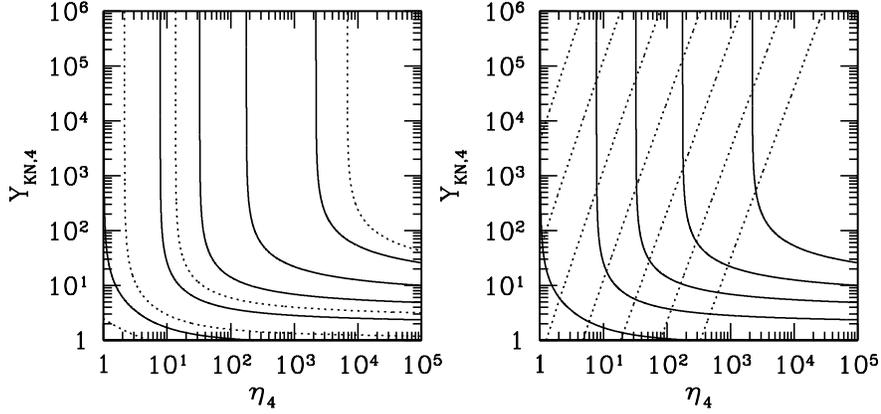}
\end{center}
\caption{ \small{ {\it Left:} Solid contour lines represent constant $\alpha$ (spectral 
slope at 10 keV), in the plane of the Compton-$Y$ parameter for electrons radiating at
10 keV, $Y_{KN,4}$, versus $\eta_4$, which indicates how deep these electrons 
are in the KN regime. From the upper right to the lower left corner, 
$\alpha = -0.1, -0.12, -0.15, -0.2,-0.3$.
The short-dashed lines are contour lines of the power-law index of electron energy 
distribution, $p_1$.  From the upper right to the lower left corner, $p_1=1.1,1.3,1.5,1.7$. 
{\it Right:} Same contour solid lines of $\alpha$ in the $Y_{KN,4}$ versus
$\eta_4$ plane, but now the short-dashed lines represent contour lines of constant
radius of emission. From left to right, $\log(R)=11,12,13,14,15,16,17$ (in cm).  
The smallest radius is the photospheric radius, 
while the largest radius is approximately the deceleration radius (see text). Both figures are for
the parameters of GRB080916C (details can be found in the next section).}}
\label{fig1} 
\end{figure*}

In this section, we determine the values of $\eta_4$ and $Y_{KN,4}$ that are
required in order to obtain the low energy spectral index close to zero at 10 keV. 
The idea is simple: we scan all possible combinations of $\eta_4$ and $Y_{KN,4}$
and determine the power-law index of the electron distribution, $p_1$, which 
can be obtained with (\ref{eq2:p_1}).  Next, we use this power-law to find the 
observed spectrum at 10 keV, $\alpha$, using (\ref{eq:alpha}).  We present the results of 
our parameter search for $\eta_4$ and $Y_{KN,4}$ that yield a certain desired
value for $p_1$ and $\alpha$ (Fig. \ref{fig1}: Left panel).


For a given $\eta_4$ and $Y_{KN,4}$ we can determine the radius, $R$, at which the 
emission is produced.  Since $Y_{KN,4}$ is defined as $Y_{KN,4} = t'_{syn}/t'_{IC,KN}$, we can 
use (\ref{t_syn_gamma_4}) and (\ref{IC_KN}) to determine $R$ as a function of 
$Y_{KN,4}$ and $\eta_4$.  We find 

\begin{equation}
R_{15} \approx 3 \frac{\eta_4^{3/2} \Gamma_3^2 L_{53}^{1/2}}{Y_{KN,4}^{1/2} (1+z) \epsilon_p^2}.
\end{equation}
We can now calculate $R$ as a function of $\eta_4$ and $Y_{KN,4}$ to find
the range of radii where synchrotron radiation should be produced for each
value of $\alpha$.  In the $Y_{KN,4}$ 
versus $\eta_4$ plane, lines of constant radius are approximately given 
by $Y_{KN,4} \propto \eta_4^3$ (Fig. \ref{fig1}: Right panel).

It is important to mention that our analytical results use a simplified 
approximation for the KN cross section in (\ref{IC_KN}), which is that 
$\sigma_{KN} \approx \sigma_T/\eta$. However, results shown in all figures 
use the full KN cross section $\sigma_{KN} = \sigma_T f(\eta)$, so 
that the radius is $R\propto \eta^2 (f(\eta))^{1/2}$, instead of 
$R \propto \eta^{3/2}$; analytical results are within a 
factor $\sim 2$ of numerical values.

\subsection{Constraint on the radius of emission}

The first constraint that we can place on the $Y_{KN,4}$ versus 
$\eta_4$ parameter space is that the cooling time scale
for electrons radiating at 10 keV
cannot be larger than the dynamical time scale.  After all, 
we need electrons to cool rapidly via IC scatterings in 
the KN regime to obtain $\alpha \approx 0$.
This constraint can be obtained in the following way.

The electron cooling time in the jet comoving frame for 
electrons of LF $\gamma_4$ is given by 
$t'_{cool} = (1/t'_{syn} + 1/t'_{IC,KN})^{-1}$.  We use
(\ref{t_syn}) and (\ref{IC_KN}) to determine $t'_{cool}$.
We impose the constraint, as discussed above, that 
$t'_{cool} < t'_{dyn}$, and with it, we find an upper 
limit on the radius of emission, $R_{cool}$, and we can use it 
to constraint our $Y_{KN,4}$ versus $\eta_4$ parameter space.
An analytical estimate of this maximum radius is provided below.

The cooling of electrons of LF $\gamma_4$ is dominated 
by the IC cooling in the KN regime.  Therefore, we can 
approximate $t'_{cool} \approx t'_{IC,KN}$ and then 
set $t'_{cool} < t'_{dyn}$, and use (\ref{IC_KN}) 
and (\ref{t_dyn}) to find

\begin{equation} \label{r_max}
R_{cool,15} \approx 220 \frac{L_{53}}{\Gamma_3^2 \epsilon_p}.
\end{equation}
Another upper limit on the radius is given by the 
radius at which the external forward shock sets in, 
that is, the deceleration radius, $R_{dec}$.
This radius is a function of the total blast wave energy, 
$E$, the circum-stellar density, $n$, which 
we assume is a constant, and the bulk LF (see, e.g., 
Sari, Piran \& Narayan 1998)

\begin{equation} \label{r_dec}
R_{dec,15} = 130 E_{55}^{1/3} n_0^{-1/3} \Gamma_3^{-2/3}.
\end{equation}
The true upper limit will be given by the minimum 
of $R_{cool}$ and $R_{dec}$.  

There is also a lower limit on the radius, given by the photospheric radius,
\begin{equation}
 R_{ph} \approx
    {L\sigma_T \over 8\pi m_p c^3 \Gamma^3} \approx (5.5\times 10^{10} 
  {\rm cm})\, L_{53}\,\Gamma_3^{-3}.
     \label{rp}
\end{equation}
Therefore, the radius of emission should lie between 
$R_{ph}$ and the minimum of $R_{cool}$ and $R_{dec}$.

\subsection{Constraint on SSC component}

The last constraint that we place on the $Y_{KN,4}$ versus $\eta_4$
parameter space is that the total luminosity in the 
synchrotron-self-Compton (SSC) component at $\sim 1$ GeV, $L_{IC}$, should 
not exceed the synchrotron luminosity, $L_{syn}$, for consistency with the 
{\it Fermi} data (Abdo et al. 2009), that is $L_{IC}/L_{syn}<1$.  The ratio 
of luminosities is given by $L_{IC}/L_{syn} \approx Y_{KN,i}$, where $Y_{KN,i}$
is the Compton-$Y$ parameter in the KN regime of electrons radiating 
at $\gamma_i$ (Nakar et al. 2009). The relationship between $Y_{KN,i}$ and 
$Y_{KN,4}$, is approximately given by

\begin{equation}
Y_{KN,i} = Y_{KN,4} \frac{f'_{syn}(<\nu'_{KN}(\gamma_i))}{f'_{syn}(<\nu'_{KN}(\gamma_4))},
\end{equation}
where $f'_{syn}(<\nu'_{KN}(\gamma_e))$ is the synchrotron flux below the KN frequency, 
$\nu'_{KN}(\gamma_e)= m_e c^2/(h \gamma_e)$, both quantities in the jet comoving frame.
The ratio of these comoving synchrotron fluxes is given by 
$\gamma_4/\gamma_i=(10$ keV$/\nu_p)^{1/2}$, therefore 

\begin{equation} \label{Y_{KN,i}}
Y_{KN,i} = Y_{KN,4} \left(\frac{10 \rm keV}{\nu_p}\right)^{1/2}.
\end{equation}

We have found that the analytical calculation of Compton-$Y$ can 
overestimate its true value by up to a factor of $\sim 10$
(Barniol Duran \& Kumar 2011, see, also, Nakar et al. 2009).
If we restrict our parameter space to $L_{IC}/L_{syn} < 1$, to avoid conflict with
{\it Fermi} high energy observations, then these last two considerations
translate to a conservative constraint on $Y_{KN,i}$ given by $Y_{KN,i} \lae 10$.
Solutions that have $Y_{KN,i}$ larger than this limit violate 
{\it Fermi} observations and are ruled out.

To summarize, we calculate the power-law index of the electron energy distribution 
function, $p_1$, as a function of $Y_{KN,4}$ and $\eta_4$ using 
(\ref{eq2:p_1}).  With $p_1$ and equation (\ref{eq:alpha}), we can determine the lower 
energy spectral index at 10 keV, $\alpha$. The 2-D space ($Y_{KN,4}, \eta_4$)
can be constrained by calculating the radius of emission that 
corresponds to each point in the $Y_{KN,4}$--$\eta_4$ plane.  The radius 
of emission should not be smaller than the photospheric radius 
nor larger than the minimum of the deceleration radius (eq. \ref{r_dec}),
and the radius at which the dynamical and cooling time scales are equal, 
(eq. \ref{r_max}). We can further constrain the parameter space 
by ensuring that the amount of energy in the SSC component is not excessive. 

\section{Application to GRB data}

\begin{figure*}
\begin{center}
\includegraphics[width=12cm, angle = 0, clip=true, viewport=.0in .0in 8in 4.5in]{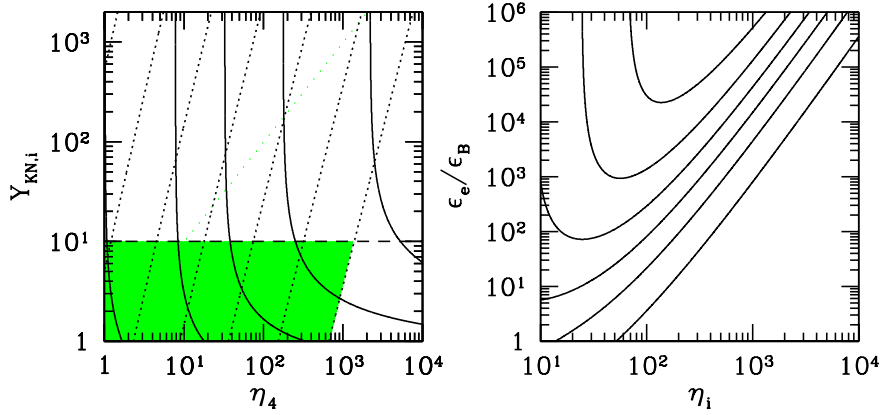}
\end{center}
\caption{ \small{ {\it Left:} The solid lines are contours of constant 
$\alpha$ (from the upper right to the lower left corner $\alpha = -0.1,-0.12,-0.15,
-0.2,-0.3$), while the 
short-dashed lines correspond to constant $R$ value in this 2-D plane; $R=10^{11}$cm
for the left most line, and the subsequent ones correspond to $10^{12}$, 10$^{13}$,
10$^{14}$, 10$^{15}$, 10$^{16}$ and $10^{17}$ cm (the largest radius is approximately
the deceleration radius). Note that we plot $Y_{KN,i}$ instead of $Y_{KN,4}$ 
(which differ only by a constant factor).
The maximum allowed value of $Y_{KN,i} \approx 10$, so as
to avoid a SSC component excess in the {\it Fermi} high energy observations, 
is plotted as a long-dashed line (the factor of 10 allows for an
overestimation in the analytical result). The allowed region is the  
region below the long-dashed line and to the left of the maximum radius, which
has been shaded. The maximum value of $\alpha^{max}=-0.11$ is obtained 
at the intersection of the maximum allowed radius and the maximum $Y_{KN,i}$.
It is clear that $\alpha \gae -0.1$ is impossible to reach for the data 
of GRB080916C. 
{\it Right:} The solid lines show $\alpha$ in the $\epsilon_e/\epsilon_B$
versus $\eta_i$ plane.  $\epsilon_e$ and $\epsilon_B$ are the fractions
of energy in electrons and magnetic field, respectively, while $\eta_i$
indicates how deep in the KN regime are the electrons radiating at the
peak frequency.
From top to bottom, $\alpha=-0.20,-0.25,-0.30,-0.35,-0.40,-0.45$. 
The data in these panels are for GRB080916C, however, the plot for
an average long-duration GRB is almost identical. }}
\label{fig2} 
\end{figure*}

We present the allowed $Y_{KN,4}$--$\eta_4$ parameter
space for two GRBs. We chose a very energetic {\it Fermi} GRB, GRB080916C, 
and another more ``standard'' GRB, which we call GRB$\natural$.
The parameters for GRB080916C are $\epsilon_p=7.5$, 
$z=4.3$, $L_{53}\approx 1$, $\Gamma=10^3$ and, for most of the duration of the 
prompt emission of this GRB, $\alpha=-0.02 \pm 0.02$ (Abdo et al. 
2009).  To find the deceleration radius, $R_{dec}$, we choose 
$E_{55}\approx 3$ (Kumar \& Barniol Duran 2010) and assume $n_0 =1$, although
the dependence on $E$ and $n$ is weak.  We choose the parameters for GRB$\natural$ to be 
more typical values of a GRB:  $\epsilon_p=3$, 
$z=2$, $L_{53}\approx 10^{-2}$, $\Gamma = 300$, $E_{55} \approx 10^{-2}$,
$n_0=1$ and, for most of the duration of the 
prompt emission of this GRB, $\alpha = 0$ (Preece et al. 2000).  Following 
the prescription at the end of last section, we present the 
results of the allowed parameter space in Fig. \ref{fig2} (Left panel).
Note that we present the parameter space now as a function of $Y_{KN,i}$, 
instead of $Y_{KN,4}$ (which is a better indicator of $L_{IC}/L_{syn}$), 
however, they are related by a constant factor, eq. (\ref{Y_{KN,i}}).


As can been seen, $\alpha \rightarrow 0$ as we move into a region 
where $\eta_4$ and $Y_{KN,i}$ (and consequently, $Y_{KN,4}$) both become larger. 
What is the maximum possible lower energy spectral index at 10 keV 
($\alpha^{max}$) that can obtained? It is given by the value where
 the maximum of $Y_{KN,i}$ intersects with the maximum allowed 
radius (see Fig. \ref{fig2}: Left panel).  This intersection 
gives us the maximum allowed value for $\alpha$ ($\alpha^{max}$) which is
found to be $-0.1$ (Table \ref{table1}). Therefore, any of the GRB 
gamma-ray radiation for which $\alpha>-0.1$ cannot be produced by the
synchrotron process in shock heated plasma, where electrons are only 
accelerated when they cross the shock front and are scattered 
back to the other side (this however, is not the case if electrons
are continuously accelerated while they are traveling downstream
or upstream; we do not consider this scenario in this paper).

In Table \ref{table1}, we also present the maximum value of $\alpha$ obtained 
if: (i) We decrease the value of the peak energy as observed during the 
prompt phase (Abdo et al. 2009), and/or (ii) We decrease the value of the LF of the source, 
as suggested by several groups (see, e.g., Zou et al. 2011, Hasco\"et et al. 2011).

The luminosity carried by the magnetic field, $L_B$, 
as measured by a lab frame observer is 

\begin{equation} 
L_B=\frac{B^2\Gamma^2}{8\pi}4 \pi R^2 c = (1.1\times10^{52} \rm erg) R_{15}^2 (\epsilon_p/\eta_4)^4 \Gamma_3^{-4}(1+z)^2,
\end{equation}
where we made use of (\ref{B_field}). Therefore, the fraction of energy carried
 by the magnetic field is

\begin{equation} \label{B_flux}
\epsilon_B=\frac{L_B}{L}=0.1 (\epsilon_p/\eta_4)^4 L_{53}^{-1} \left(\frac{R_{15} (1+z)}{\Gamma_3^{2}}\right)^2.
\end{equation}
We present the value of $\epsilon_B$ for the $\alpha^{max}$ case (Table
\ref{table1}); $\epsilon_B$ is found to be very small, in the range 
$\sim 10^{-6}-10^{-4}$. We also provide in Table 1 the value of the LF 
of electrons that radiate at the peak of the spectrum ($\gamma_i$) 
in order for the low energy spectral slope to be $\alpha^{max}$; 
we find $\gamma_i \ge 10^6$.

\begin{table}
\begin{center}
\begin{tabular}{ccccccc}
\hline
GRB & $\Gamma_3$ & $\epsilon_p$ & $\eta_4$ & $\gamma_i$ & $\epsilon_B$ &
$\alpha^{max}$ \\
    &            &              &          &($\times 10^{6}$)& ($\times
10^{-5}$)& \\ 
\hline 
080916C  & 1    & 7.5 & 1300  & 2  & 3   & -0.11 \, (-0.14)\\
         & 0.3  & 7.5 & 16500 & 6  & 0.3 & -0.10 \, (-0.14)\\
         & 1    & 2.5 & 400   & 1  & 10  & -0.14 \, (-0.20)\\
         & 0.3  & 2.5 & 3400  & 2  & 2   & -0.13 \, (-0.20)\\
\hline
$\natural$  & 0.3 & 3 & 2500  & 2  & 2   & -0.11 \, (-0.17)\\
            & 0.1 & 3 & 28000 & 7  & 0.2 & -0.10 \, (-0.17)\\
            & 0.3 & 1 & 900   & 1  & 7   & -0.15 \, (-0.25)\\
            & 0.1 & 1 & 5700  & 2  & 1   & -0.14 \, (-0.25)\\

\hline
\end{tabular}
\end{center}
\caption{{\small We present the maximum possible spectral slope,
$\alpha^{max}$, for two GRBs: GRB080916C and GRB$\natural$.  
$\alpha^{max}$ is the value of $\alpha$ obtained, where the 
maximum $Y_{KN,i}$ intersects with the maximum radius (see Fig. 
\ref{fig2}: Left panel).
We present results for different values of the GRB-jet LF and the observed 
spectral peak ($\epsilon_p$; see eq. 1 for definition).  
In addition, we also present the values for $\eta_4$, $\gamma_i$ and $\epsilon_B$, 
where the spectral slope is maximum.  The value of $\alpha$ does not 
reach the observed value $\alpha=-0.02 \pm 0.02$ (Abdo et al. 2009) for
GRB080916C, nor $\alpha = 0$ for GRB$\natural$ for any combination of
parameters. At $\alpha^{max}$ we find extreme values for 
the LF of electrons radiating at the peak and for the energy fraction in the 
magnetic field. The values of $\alpha^{max}$ where we assume a variability 
time scale of $\delta t = 0.1$ s are in parenthesis (see next Section).}}
\label{table1}
\end{table}

If we take the observer frame variability time scale of the gamma-ray
light curve to be 

\begin{equation} \label{r_is}
\delta t = \frac{R (1+z)}{2 c \Gamma^2},
\end{equation}
as expected in the internal shock model, then in order for the 
low energy spectral slope to be $\alpha^{max}$, we find 
$\delta t \ge 10$ s, since the emission is produced at 
(or very close to) the deceleration radius.  This is very long 
compared with the observed time scale of $0.1$ s or less.

We can calculate the low energy spectral index as a function of 
$\epsilon_e/\epsilon_B$,  where $\epsilon_e$ is the fraction
of energy in electrons.  This will allow us to compare our results 
with previous work that use this ratio.  To do this, we use (Ando et al. 2008)

\begin{equation} \label{Y_{KN,i,2}}
Y_{KN,i} = \sqrt{\frac{\epsilon_e/\epsilon_B}{\eta_i}},
\end{equation} 
where $\eta_i$ parameterizes how deep electrons of $\gamma_i$ 
are in the KN regime, analogous to (\ref{gamma_4}), and it is 
related to $\eta_4$ as  
$\eta_i/\eta_4 = \gamma_i/\gamma_4=(\nu_p/10$ keV$)^{1/2}$.  Using
(\ref{Y_{KN,i}}) and (\ref{Y_{KN,i,2}}), we find $\epsilon_e/\epsilon_B$
as a function of $\eta_4$ and $Y_{KN,4}$ and calculate 
$p_1$ using equation (\ref{eq2:p_1}), and $\alpha$ from equation (\ref{eq:alpha}).
The result is shown in Fig. \ref{fig2} (Right panel).

\section{Discussion} 

Most GRBs have low energy spectral index $\alpha$ between $0$ and $-0.1$; 
$f_\nu\propto \nu^\alpha$. For $\alpha > -0.1$, it is required that 
 the power-law index for electron energy distribution function, 
$p_1\equiv|\frac{d \ln n_e}{d \ln \gamma_e}|$, should be less than 1.15 at 
$\gamma_e$ corresponding to 10 keV synchrotron photons (in observer frame).
For this, two conditions must be satisfied. 
1. The Compton-$Y$ parameter for electrons radiating at 10 keV
should be $Y_{KN,4} \gae 20$ (including KN effect), and 2. The LF of electrons 
radiating at 10 keV should be $\gae 2000 \Gamma \epsilon_p^{-1}$, where 
$\epsilon_p=\frac{h\nu_p(1+z)}{m_ec^2}$, 
that is, $\eta_4 \gae 2000$ (see Fig. \ref{fig1}).  These conditions 
apply both for the data of GRB080916C (a highly energetic explosion)
and also for an average long duration burst.

Electron index changes sharply, from being close to 1 (when IC in
the KN regime dominates) at 10 keV to $>$2 at the peak of the observed
spectrum at $\sim$ 1 MeV, that is, the electrons distribution index increases
from $\sim$1 to $>$2, when the electron LF increases by a factor $\sim10$. In this
case, the spectral index for synchrotron radiation at 10 keV is not 
given by $-(p_1-1)/2$ (as mentioned
in the Introduction); in fact, it is significantly smaller. For instance, when 
$p_1 = 1.02$ (for electrons radiating at 10 keV), 
$\frac{d \ln f_{\nu}}{d \ln \nu}= - 0.057$ and not $-0.01$ as naively expected;
see (\ref{eq:alpha}).  

Another consequence of the requirement that $\eta_4 \gae 2000$ is the that energy
fraction in magnetic field, $\epsilon_B$, is rather small
$\epsilon_B \lae 10^{-9} R_{15}^2$ (GRB080916C) and 
$\epsilon_B \lae 10^{-7} R_{15}^2$ (for an average burst); 
see (\ref{B_flux}).  In addition, the LF of electrons radiating 
at the peak of the spectrum ($\sim 1$ MeV) is $\sim 10$ times larger
than that of electrons radiating at 10 keV, therefore, 
$\gamma_i \gae 2 \times 10^4 \Gamma \epsilon_p^{-1}$, 
which is $\gamma_i \gae 10^6$ almost independent of GRB energy (Table 1).
Both values of $\epsilon_B$ and $\gamma_i$ are extreme and their 
implications will be discussed in the Conclusions.

 However, we 
can also ask: what is $\alpha$ for a reasonable set of 
parameters? In this case, ``reasonable'' means two things:
(1) The radius of emission of the prompt emission should be 
between (i) the photospheric radius and (ii) the radius where 
electrons producing 10 keV synchrotron photons cool on
a time scale shorter than the dynamical time via IC scatterings in 
the KN regime, or the deceleration radius, whichever is smaller, 
and (2) The energy in the IC component
should not be very large, so that the GRB spectrum 
does not show an IC bump at $\sim 1$ GeV, as the {\it Fermi} satellite
sees no sign for such an excess.
We have calculated the maximum value of $\alpha$ that 
can be obtained for a highly energetic Fermi burst, GRB080916C, and also
for an average GRB and the results are presented in Table \ref{table1}. We find
that the maximum value of $\alpha$ is $\alpha = -0.1$.  For 
this value, the same consequences as discussed above apply, 
and can be found in Table  \ref{table1}. Namely, that 
$\eta_4 \sim 10^3$, and this implies a very large value of $\gamma_i$
and an extremely small value of $\epsilon_B$.  Moreover, 
the maximum value of $\alpha$ occurs at a very large
radius, close to the deceleration radius, which,  
will have problems producing variable light curves with 
$\delta t$ smaller than a few seconds.   
   
Conversely, we can also fix the observed variability time scale 
of the gamma-ray light curve to be $\delta t =0.1$ s and 
determine the radius of emission with (\ref{r_is}), 
as expected in the internal shock model.
We can determine the maximum value of $\alpha$ at 
this radius and its value is presented in parenthesis 
in Table \ref{table1}.  Notice that this value of $\alpha$ is 
even further away from the observed value. However, 
at this radius, the values of $\epsilon_B \sim 10^{-4}-10^{-3}$
and $\gamma_i \sim 10^3-10^5$ are less extreme 
and in marginal agreement with the internal shock model.  
Nevertheless, this further accentuates  
the fact that synchrotron emission, 
in which electrons cool mainly via IC in the KN 
regime, in the context of the internal shock model, 
cannot explain the observed $\alpha \sim 0$ spectrum
of most GRBs.

Expressing our analytical results as a function of $\epsilon_e/\epsilon_B$ 
allows us to compare them with previous numerical work.
Nakar et al. (2009) (see, also, Fan 2010) 
have found numerically that with 
$\epsilon_e/\epsilon_B=100$ ($10^4$), $\alpha$ cannot 
exceed $\approx -0.3$ ($-0.2$). In this work, we 
analytically confirm their results (see right panel of 
Fig. \ref{fig2}).  Daigne et al. (2011) have also found
numerically  
that $\alpha = 0$ is possible in the case 
when electrons are injected instantaneously.  However, 
in Nakar et al. (2009) and the present work, we have 
assumed that electrons are injected regularly over a 
time scale comparable to the dynamical time scale. 
Nevertheless, Daigne et al. (2011) present numerical 
results for our scenario in their fig. 2 (bottom left panel) 
for $\eta_i=100$, which agree with our analytical calculation.

Our work and the work by Daigne et al. (2011), as mentioned 
before, differ mainly on the chosen time scale at which electrons 
are injected to the shock, $t_{injec}$.  We take $t_{injec} \sim t_{dyn}$, 
where $t_{dyn}$ is the dynamical time scale, whereas Daigne et al. take 
$t_{injec} \ll t_{dyn}$.  We find a stricter limit on the allowed value of 
$\alpha$: $\alpha \lae -0.2$, whereas Daigne et al. find that solutions with 
$\alpha = 0$ can be reached (see their fig. 2).  These results are not in contradiction, since
both studies probe two different phases found in the internal shock model.  
The first phase ($t_{injec} \sim t_{dyn}$) has electrons continuously being 
injected to the shock as it crosses the shell.  The second phase 
($t_{injec} \ll t_{dyn}$) corresponds to the case where the shock has already 
traversed the shell and electrons cool as the shell adiabatically expands.  
Most of the available energy is dissipated in the first phase; for this reason 
we have chosen this particular scenario in this paper, which is the common 
practice in studies of the internal shock model (see, e.g., Piran 1999).  

However, when calculating synthetic GRB light curves, Daigne et al. (2011) do follow the
dynamics of the shock crossing numerically, and consider a large number
of discretized shells on the dynamical time scale. In each collision the electron 
injection occurs instantaneously, $t_{injec} \ll t_{dyn}$, but the electron injection 
process over the full simulation is comparable to the dynamical timescale. 
In this case, when including IC cooling in the KN regime (see their fig. 9), 
they find $\alpha \lae -0.1$. The difference in our results appears because the IC
scatterings in their work do not occur between the same photon and electron distributions
we have considered: In Daigne et al. (2011) the scatterings between photons emitted 
in a shocked region and electrons or photons present 
in a subsequent shocked region were not considered (see Bo\v{s}njak et al. 2009).  
This affects the cooling of the electrons via IC, allowing them to 
reach spectra closer to $\alpha = 0$.

\section{Conclusions}

In this work we have investigated the possibility that 
the observed low energy GRB prompt spectrum, which is 
$f_{\nu} \sim \nu^0$ (below the peak) for a good fraction
of all long duration GRBs (Preece et al. 2000; Kaneko et al. 
2006; P\'{e}langeon et al. 2008; Krimm et al. 2009; Ghirlanda et al. 2010),
is due to synchrotron radiation from electrons that cool mainly via the IC 
mechanism in the KN regime (Derishev et al. 2003,
Bo\v{s}njak et al. 2009, Nakar et al. 2009, Wang et
al. 2009, Fan 2010, Daigne et al. 2011). 

We present an analytical method to determine
the power-law index of the electron energy 
distribution function, $p_1$, that cools via IC
cooling in the KN regime as a function of 
two parameters: $\eta$, which is a measure of how 
deep electrons of interest are in the KN regime, 
and $Y_{KN}$, which is the Compton-$Y$ parameter 
(including KN corrections) for these electrons.  We have calculated 
the observed low energy spectral index for synchrotron radiation, $\alpha$, 
as a function of these two parameters as well as the power-law
index of the electron energy distribution above $\gamma_i$
($\gamma_i$ corresponds to the LF of electrons radiating at 
the peak). We find that $\alpha$ is not simply given by 
$-(p_1-1)/2$ as naively expected, but it is smaller, which
makes it very difficult to explain the observed value of 
$\alpha\approx 0$ for a good fraction of GRBs.

We find that $\alpha>-0.1$ cannot be obtained for parameters
relevant for GRBs, if the radiation mechanism is the synchrotron 
process and electrons are accelerated in a relativistic shock,
where electrons are only accelerated when they cross the 
shock front and are scattered back to the other side.
Therefore, the $\gamma$-ray radiation from a significant fraction 
of long duration GRBs that have low energy spectral index larger 
than $-0.1$ cannot be accounted for by this mechanism.

Even $\alpha \approx -0.1$ faces severe difficulties. 
The large radius for generation of $\gamma$-rays 
is in conflict with the short variability time ($\lae$
0.1 s) of prompt GRB light curve. Moreover, the 
energy in the magnetic field must be extremely small, 
$\epsilon_B \sim 10^{-6}-10^{-4}$, and $\gamma_i \ge 10^6$ 
for the mechanism to be able to able to harden 
the spectral slope from $\alpha=-0.5$ to $\sim -0.1$ (Table 1).

It is unlikely that the energy fraction in the 
magnetic field will be so small ($\epsilon_B<10^{-4}$) in internal
shocks. If the central engine of GRBs is powered by accretion onto
a black hole, we expect $\epsilon_B\sim 1$\% as magnetic fields of 
such a strength are likely produced in the accretion disk by the
Balbus-Hawley mechanism (Hawley, Gammie \& Balbus 1996); 
for a magnetar based central engine this small $\epsilon_B$ is even more surprising.

For the typical thermal LF of electrons to
be large, $\gamma_i\gae10^6$, in internal shocks where
shells collide with a relative LF of a few to 10, it is required 
that approximately only 1 in $\sim10^3$ electrons are 
accelerated when they cross the shock-front but they receive 
$\sim10$\% of the total energy. This is in contradiction with the
numerical PIC simulations of Sironi \& Spitkovsky (2011). 
Moreover, the $\sim 99.9$\% of 
electrons which are not accelerated
have a thermal LF of a few thousand due to their
interaction with protons (Sironi \& Spitkovsky 2011), and these 
electrons produce a significant IC bump in the spectrum at $\sim$ 100 MeV 
which is not seen for any bursts. The SSC flux of these electrons at 
100 MeV will be very large: about a factor 
of 10 larger than the observed flux. 

All these difficulties suggest that the synchrotron radiation mechanism 
in internal shocks does not provide a self-consistent solution when 
the low-energy spectral index for GRBs is larger than about $-0.2$.

\section*{Acknowledgments}
RBD dedicates this work to Adolfo Barniol, and
thanks Jessa Barniol for her support during the writing of this
manuscript.  RBD thanks Ehud Nakar, Tsvi Piran and Paz Beniamini 
for useful discussions. This work has been funded in part by NSF grant
ast-0909110. ZB acknowledges the French Space Agency (CNES) for 
financial support.


\end{document}